\documentclass[12pt]{article}

\usepackage{algorithm,algpseudocode}
\usepackage{amscd,amsfonts,amsopn,amssymb,amstext}
\usepackage{appendix}
\usepackage{booktabs}
\usepackage{bbm,bm}
\usepackage[centertags]{amsmath}
\usepackage{color}
\usepackage{fullpage}
\usepackage{graphicx,graphics,psfrag}
\usepackage{indentfirst}
\usepackage{latexsym,enumerate}
\usepackage{lscape}
\usepackage{multirow}
\usepackage{natbib}
\usepackage{rotating}
\usepackage{setspace}
\usepackage{srcltx}
\usepackage{subfigure}
\usepackage[T1]{fontenc}
\usepackage{threeparttable}
\usepackage{times}
\usepackage{url}
\usepackage{verbatim}

\makeatletter
\renewcommand\@biblabel[1]{#1.}
\makeatother

\newcommand{\qed}{\rule{0.5em}{1.5ex}}

\title{\Large \bf{A field- and time-normalized Bayesian approach to measuring the impact of a publication}}

\author{\normalsize
{\bf Emilio G\'omez--D\'eniz, Pablo Dorta--Gonz\'alez\thanks{P. Dorta--Gonz\'alez, Department of Quantitative Methods in Economics,
University of Las Palmas de Gran Canaria, 35017 Las Palmas de Gran Canaria, Spain. E-mail: \texttt{pablo.dorta@ulpgc.es}}, }\\
{\small Department of Quantitative Methods in Economics and TiDES Institute,}\\[-0.2cm]
{\small University of Las Palmas de Gran Canaria, Spain}\\[-0.15cm]
{\small http://orcid.org/0000-0002-5072-7908 (EGD) }\\[-0.20cm]
{\small http://orcid.org/0000-0003-0494-2903(PDG)}
}

\date{}

\def \E{{\rm I\kern -2.2pt  E}}

\begin{document}

\maketitle

\vspace{-0.5cm}

\begin{abstract}\noindent  Measuring the impact of a publication in a fair way is a significant challenge in bibliometrics, as it must not introduce biases between fields and should enable comparison of the impact of publications from different years. In this paper, we propose a Bayesian approach to tackle this problem, motivated by empirical data demonstrating heterogeneity in citation distributions. The approach uses the a priori distribution of citations in each field to estimate the expected a posteriori distribution in that field. This distribution is then employed to normalize the citations received by a publication in that field. Our main contribution is the Bayesian Impact Score, a measure of the impact of a publication. This score is increasing and concave with the number of citations received and decreasing and convex with the age of the publication. This means that the marginal score of an additional citation decreases as the cumulative number of citations increases and increases as the time since publication of the document grows. Finally, we present an empirical application of our approach in eight subject categories using the Scopus database and a comparison with the normalized impact indicator Field Citation Ratio from the Dimensions AI database.

\vspace{-0.5cm}
\paragraph{Keywords:}Normalized citation impact, field normalization, time normalization, Bayesian score, citation obsolescence, citation potential, citation density.

\paragraph{Mathematics Subject Classification (2020):} 62E10, 62P25.
\end{abstract}

\subsubsection*{Funding:}
Emilio G\'omez-D\'eniz was partially funded by grant PID2021-127989OB-I00 (Ministerio de
Econom\'ia y Competitividad, Spain).

\section{Introduction}
There is a large literature on field-normalized citation counting. An overview of this topic can be found in \cite{bornmannandmarx_2015}, and \cite{waltman_2016}. However, the bibliography is much smaller in relation to time normalization. In general, for simplicity, the accumulated citations of a document are divided by the number of years since its publication. In this way, a citation is given the same value regardless of when it occurred.

The use of research publications tends to decrease as the literature ages, and authors tend to cite more recent documents while neglecting older ones, a phenomenon known as literature obsolescence. Despite this, older papers still have had more time to accumulate citations. Modeling literature obsolescence has been studied since the 1960s, with \cite{desolla_1965} proposing a negative exponential distribution for the decline of literature use over time, while others suggest a lognormal distribution as a better measure \citetext{\citealp{Egghe1992}; \citealp{gupta_1998}}. However, due to the ever-increasing rate of scientific publication \citetext{\citealp{bornmannandmarx_2015}}, older influential papers may have had lower citation potential in the early stages of their publication compared to younger influential papers.

Both because of the phenomenon of obsolescence, which causes the density of citations to decrease as the age of the cited document increases, and because of the increasing rate of growth of the scientific corpus, citations to older documents should be given more weight than citations to more recent documents. Greater weight in the sense of greater recognition when measuring the cumulative impact of a publication.

The empirical citation data show heterogeneity in the distributions. Therefore, the Bayesian approach could be used to address the problem of field and time normalization. In this approach, the a priori distribution of citations in each field is used to estimate the expected a posteriori distribution in that field. This a posteriori distribution is then used to normalize the citations received by a publication in that field.

Regarding the impact of publications, \cite{perezetal_2013} propose a Bayesian approach that utilizes a weighting scheme based on previous impact factors of specific journals. This method aims to obtain a reliable and consistent version of the actual impact factor. The approach depends on selecting a probability distribution for the citation process and a prior distribution over the parameters. They suggest a Poisson-Gamma family, which extends the negative binomial distribution and is suitable for analyzing over-dispersed data, with the mean being the impact factor of the journal in the previous year.

However, the Journal Impact Factor favours journals that concentrate a large proportion of their citations in the first few years after publication, i.e. journals in fields with high obsolescence. Thus, a proposal that gives an increasing marginal score over time could benefit fields with less obsolescence.

This paper adopts a Bayesian approach to tackle the issues of field and time normalization and proposes a Bayesian Impact Score as a measure of a publication's impact. The score is characterized as increasing and concave with the number of citations received and decreasing and convex with the age of the publication. This implies that the additional value of a citation decreases as the total number of citations increases, while it increases with the time since the publication. Therefore, the impact score assigns less weight to citations received in the early years after publication and gradually reduces their influence as the total number of citations grows.

The paper is organized into several sections. Section 2 presents the theoretical framework, while section 3 introduces the empirical data. Section 4 describes the model used in the study. Section 5 introduces the Bayesian Impact Score, discussing its properties and elasticity. The paper then presents in section 6 the results of numerical experiments conducted on eight subject categories from the Scopus database and a comparison with the normalized impact indicator Field Citation Ratio from the Dimensions AI database. Finally, section 7 provides conclusions based on the study's findings.

\section{Theoretical framework for field-normalized citation counting}
Bibliometrics has a long-standing tradition of normalization, with literature reviews on this topic available in \cite{bornmannandmarx_2015} and \cite{waltman_2016}. The problem of field-specific differences in citation counts arises in the evaluation of research institutions. Interdisciplinary research institutes, in particular, typically draw scholars from diverse disciplinary backgrounds \citetext{\citealp{wagneretal_2011}}, making it challenging to compare citation counts directly.

One of the fundamental principles of citation analysis is that the citation counts of publications from different fields should not be compared directly, as there are significant differences in citation density. Citation density refers to the average number of citations per publication and is responsible for the bias effect called "citation potential," a term introduced by \cite{garfield_1979} based on the average number of references cited in a publication. For instance, the biomedical field often has long reference lists with over fifty references, while mathematics typically has short lists with less than twenty references \citetext{\citealp{dortanddorta_2013a}}.

Moreover, even within the same field, comparing the citation counts of publications from different years directly, even after dividing by the age of publication, is not recommended due to significant differences in the average number of citations per year \citetext{\citealp{dortanddorta_2013b}}. These variations result from differences in citation habits, which influence both the number of citations and the likelihood of being cited \citetext{\citealp{leydesdorffandbornmann_2011}; \citealp{zittandsmall_2008}}.

However, for practical purposes, it is necessary to compare publications from different fields, years, or document types. In order to facilitate such comparisons, normalized citation impact indicators have been proposed.

\subsection{Normalized indicators based on average number of citations}
Normalized indicators rely on the concept of the expected number of citations for a publication, which is determined by calculating the average number of citations of all publications in the same field, from the same year, and of the same document type. Field normalization is traditionally based on a classification system, such as the WoS journal subject categories, although there are other options available \citetext{\citealp{bornmannandwohlrabe_2019}}. Under this method, each publication is allocated to one or more fields, and its citation impact is evaluated in comparison to other publications within the same field.

The relative citation rate (RCR) was among the earliest techniques developed for normalizing citations across fields. \cite{schubertandbraun_1986} and \cite{vinkler_1986} were the originators of this method. They determined the average citation rate for a specific field or journal and used it as a reference score to normalize papers published within the same field or journal. This was accomplished by dividing the citation counts of each paper by the reference score. The Field-Weighted Citation Impact (FWCI) is a variant of this indicator, although it is highly correlated with the RCR \citetext{\citealp{purkayasthaetal_2019}}. However, the use of the arithmetic mean in RCR normalization has been criticized because it is not suitable for skewed distributions with a long tail as it is sensitive to outliers \citetext{\citealp{glanzelandmoed_2013, vanraan_2019}}.

A variant of the RCR can be obtained for a set of publications in two ways. One way is by calculating the average of the normalized citation scores of the publications in the set. The other way is by computing the ratio of the total number of actual citations to the expected number of citations for the same set of publications. There is no consensus in the literature as to which of the two approaches is better. However, most researchers seem to prefer the first approach, which is the average of ratios, over the second approach, which is the ratio of averages. This preference is evident in the works of \cite{lundberg_2007}, \cite{opthofandleydesdorff_2010}, \cite{thelwall_2017} and \cite{vanraanetal_2010}. Nevertheless, some authors \citetext{\citealp{moed_2010}; \citealp{vinkler_2012}} prefer the average ratio approach. Empirical studies comparing the two approaches indicate that the differences are minor, especially at the level of research institutions and countries \citetext{\citealp{herranzandruizcastillo_2012}; \citealp{lariviereandgingras_2011}}.

This paper uses a basic methodology centred on the normalization of actual citation counts through the integration of expected citation counts. Currently, the three major general purpose bibliographic databases provide standardized citation indicators. Expected citation counts are defined as the average citation counts of papers within the same field and year, giving rise to the concept of CNCI (Category Normalised Citation Impact). CNCI is a metric closely related to Clarivate Analytics' InCites database. It is similar to the Field-Weighted Citation Impact (FWCI) in Scopus' SciVal database and the Field Citation Ratio (FCR) in the Dimensions AI database.

These metrics, including CNCI, FWCI and FCR, share the common goal of providing a normalized assessment of citation impact. They do this by accounting for the variation that naturally occurs across fields and years. In essence, they provide a more nuanced and equitable assessment of scholarly impact by considering the different contexts in which scholarly research is conducted.

\subsection{Normalized indicators based on percentiles}
McAllister et al. (1983) suggested using percentiles to address skewed distributions. Typically, a field-specific threshold is selected to identify highly cited publications. For example, \cite{tijssenetal_2002} consider the top 1\% and 10\% most highly cited publications, while \cite{vanleewenetal_2003} focus on the top 5\%, and \cite{gonzalezanddorta_2017} estimate the top 10\% most highly cited publications empirically. \cite{bornmannandwilliams_2020} provide a recent review of percentile measures.

However, determining the exact number of publications above a given threshold is often impractical because many publications in a field have the same number of citations. \cite{pudovkinandgarfield_2009} and \cite{leydesdorffetal_2011} propose some solutions to this issue. \cite{waltmanandscheriber_2013} review this topic, and \cite{schreiber_2013} provide an empirical comparison.

Alternatively, \cite{leydesdorffetal_2011} propose dividing publications into several classes based on percentiles of the citation distribution in a field (e.g. below the 50th percentile, between the 50th and 75th percentile, etc.), instead of distinguishing between highly cited and not highly cited publications. A similar approach is presented by \cite{glanzel_2013} and \cite{glanzeetal_2014}.

Finally, it is worth noting that older papers have had more time to accrue citations. However, as the scientific literature expands at an accelerating rate \citep{bornmannandmutz_2015}, influential papers that are older may have had a lower potential for citations in the short term than similarly influential but younger papers. Furthermore, calendar year-based normalization methods can yield imprecise results for recently published papers \citep{ioannidisetal_2016}. In a recent empirical study comparing normalized indicators, \cite{dunaiskietal_2019} found that percentile-based citation scores are less affected by field and time biases than mean-based citation scores.

\section{Empirical data}

The empirical application used the Scopus database as the primary data source. Eight subject categories (subfields) were considered, including three from the natural sciences, two from the social sciences, one from the health sciences, one from engineering and one from the humanities: Cell Biology, Economics \& Econometrics, Electrical \& Electronic Engineering, General Chemistry, General Medicine, General Physics \& Astronomy, History and Library \& Information Science.

The selection of these subfields was guided by the authors' previous experience to include disciplines with diverse characteristics. Within each subfield, one {disciplinary journal} was selected from the top 10\% of the most cited journals, as determined by the Scopus CiteScore. Then, for each selected journal, all research articles published in 2019 and catalogued in the database were collected. Finally, for each research article, the publication years of all cited references less than 150 years old were downloaded. Table \ref{tab1}, which was used in \cite{dortaandgomez_2022}, shows the index of dispersion ($\mbox{ID}=var(X)/\E(X)$) together with the mean and variance of the empirical data.

\begin{table}[htbp]

       \caption{Mean, variance and index of dispersion of the different journals studied\label{tab1}}

\begin{center}
\begin{tabular}{lcccc}\hline\hline

& Ag. Cell & Amer. Econ. Rev. & IEEE Comm. Mag. & Acc. Chem. Res.\\ \cline{2-5}

Mean & 9.21893 & 13.6897 & 4.68677 & 8.59967\\
Variance & 61.9357 & 258.575 & 49.128 & 115.578 \\
ID & 6.71832 & 18.8883 & 10.4823 & 13.4398 \\ \hline\hline

&  BMC Med.  & Adv. in Theor. \& MP & Am. Ant. & IEEE Trans. Inf. Th.\\ \cline{2-5}

Mean &  8.22895 & 17.5842 & 21.6714 & 14.6476 \\
Variance &  69.0984 & 270.698 & 467.049 & 254.451 \\
ID &  8.39699 & 15.3944 & 21.5514 & 17.3715 \\ \hline\hline

\multicolumn{5}{l}{{\scriptsize Note: Ag. Cell: Aging Cell; Amer. Econ. Rev.: American Economic Review; IEEE Comm. Mag.: IEEE Communications Magazine;}}\\
\multicolumn{5}{l}{{\scriptsize Acc. Chem. Res.: Accounts of Chemical Research; BMC Med.: BMC Medicine; }}\\
\multicolumn{5}{l}{{\scriptsize Adv. in Theor.  \& MP: Advances in Theoretical and Mathematical Physics; Am. Ant.: American Antiquity;}}\\
\multicolumn{5}{l}{{\scriptsize IEEE Trans. Inf. Th.: IEEE Transactions on Information Theory.}}
\end{tabular}
\end{center}
\end{table}

To empirically analyze the behavior of our Bayesian Score, we used the Dimensions AI database and its article-level normalized impact indicator, the Field Citation Ratio (FCR). This choice was made because this database offers Open Metrics, which provides open access to both citation counts and the FCR. This contrasts with similar normalized indicators in Web of Science (InCites) and Scopus (SciVal), where a subscription to these services is required to access such information.

The FCR serves as a metric to measure the relative citation performance of a publication compared to other articles within its field and of similar age. An FCR greater than 1 indicates an above-average citation impact as determined by the journal classification system used in the Dimensions AI database, specifically the Field of Research (FoR) subject code and publication year. This calculation is applied to all publications in Dimensions AI that are at least 2 years old and published after the year 2000.

We focused on the eight journals listed in Table \ref{tab1}. We randomly selected a simple sample of size $N=5048$ research articles from the 2014-2021 cohort, each with a calculated FCR. This represents approximately 32\% of the total articles published during these years. We downloaded the total number of citations from the database up to the year 2023. In addition, we retrieved the FCR values and calculated the corresponding Bayesian Score based on the definition given in this paper.

\section{The model}
We consider that $X_1,X_2,\dots,X_t$ are independent and identically distributed random variables with values in ${\cal X}\in{\mathbb N}=\{0,1,\dots\}$. They represent the number of citations of a journal or a collection of journals in a subject category (in general, a set of papers from the same year and as homogeneous as possible on the topic addressed) in the last $t$ years. Following to \cite{Burrell2002} we are going to assume that $X\equiv X_i$ initially follows a Poisson distribution with mean $\theta\in\Theta>0$. That is,
\begin{eqnarray}
f(x|\theta)=\Pr(X =x)=\frac{\theta ^{x}}{x!}\exp(-\theta),\quad x=0,1,\dots\label{poisson}
\end{eqnarray}

Based on empirical data, it appears that citations of papers tend to decline over time, particularly after reaching a certain peak. This trend can be represented through a Poisson distribution model. Nevertheless, the Poisson distribution assumes equidispersion, where the variance is identical to the mean, rendering it unsuitable for defining the random variable $X$. Empirical research has demonstrated that $X$ exhibits overdispersion, whereby the variance surpasses the mean. 

Additionally, it has been proposed that over-dispersion is associated with the heterogeneity within the population of subject categories. Under these circumstances, the parameter $\theta$ can be treated as a random variable that varies across distinct journals within the same subject category. This reflects the uncertainty surrounding this parameter, with its value fluctuating from one entity to another, following a probability density function. In this situation, it is assumed that the parameter adheres to a gamma distribution, characterized by a shape parameter of $\alpha>0$ and a rate parameter of $\beta>0$, i.e.
\begin{eqnarray}
\pi(\theta) = \frac{\beta^{\alpha}}{\Gamma(\alpha)} \theta^{\alpha-1}\exp(-\beta\theta),\quad \theta>0.\label{gamma}
\end{eqnarray}

Thus, the unconditional distribution results a negative binomial distribution, $X\sim NB(\alpha,1/(\beta+1))$. It is well-known that for mixed Poisson distributions the variance is always greater than the mean.

Other mixing distributions, such as the inverse Gaussian distribution, can be considered in practice, in addition to the gamma distribution. In line with this, \cite{burrell_2005} explored the use of the negative binomial and beta distributions as mixing distributions, resulting in the Waring distribution.

\section{Approaching the Bayesian Impact Score via a loss function}
The number of cites of a journal or a subject category of journals is thus specified by the random variable $X$ following the probability mass function $f(x|\theta)$ which depends on an unknown parameter $\theta$. In our case $X$ is assumed to follow a Poisson distribution. An impact score can be considered as a functional, say $F\in{\cal F}$, that assigns to $X$, the number of cites, a positive real number $F(X)$. Here ${\cal F}$ is the set of possible values of the impact score. Let now $L:X\times
{\cal F}\rightarrow \mathbb{R}^+$ be a loss function that assigns to any
$(x,F)\in \mathbb{N}\times {\cal F}$ the loss sustained, $L(x,F)$, when $X$ takes the value $x$ and the impact score of the journal is $F$. The impact score we propose can be viewed as a function that assigns to each value $x\in\mathbb{N}=\{0,1,\dots\}$ a value within
the set ${\cal F}\in\mathbb{R}^{+}$, the action space. The impact score is assumed to take only positive values. The impact score should be determined such that this expected loss is minimised. Thus,
\begin{eqnarray*}
F =\arg\min_{F} \E[L(x,F)].
\end{eqnarray*}

The quadratic loss function, $L(x,F)= (x-F)^2$, is the most commonly used loss function. Here, the unknown $F$ is the mean of the random variable $X$, i.e., $F(\theta)=\E(X|\theta)$, where $F$ is dependent on $\theta$. While other loss functions exist, the use of this loss function is often preferred due to its symmetry, which allows for fair compensation between gains and losses.

The above procedure describes the impact score for a journal when $\theta$ is known. However, the population of journals results in practise heterogeneous, among the collective of journals in the subject category for which the journal belongs. This can be seen as that $\theta$ is a particular realization of a random variable $\Theta$, with a prior distribution $\pi(\theta)$. If prior experience is not available, the corresponding impact score can be computed by minimising the risk function, i.e. by minimising $
{\E}_{\pi(\theta)}\left[L(F(\theta),F)\right]$, as $F=\E_{\theta}\left[F(\theta)\right]$, which is the mean of the unknown impact score $F(\theta)$ among all the journals in the population of journals (the subject category).

In practice, it is common to have some prior experience or information available. In this case, we can use a sample $\tilde x = (x_1, \dots, x_t)$ from the random variable $X$ to estimate the unknown impact score $F(\theta)$ by means of the Bayes estimate $F^{\ast}$, which minimises the Bayes risk. This involves minimising $\E_{\theta} \left[ L(F(\theta), \theta)\right]$, where the expectation is taken with respect to $\pi(\theta|\tilde x)$, the posterior distribution of the risk parameter $\theta$ given the sample information $\tilde x$.

Observe that given the sample $\tilde x$ from \eqref{poisson} and assumed \eqref{gamma} as the prior distribution we get that the posterior distribution of $\theta$ given $\tilde x$ is proportional to
\begin{eqnarray*}
\pi(\theta|\tilde x) \propto  \theta^{\alpha+x^+-1}\exp\left[-(\beta+t)\theta\right],
\end{eqnarray*}
where $x^+=\sum_{i=1}^{t} x_i$ is the number of cites in the sample. Therefore, this posterior distribution is again a gamma, i.e. the likelihood and prior are a conjugate pair, but with updated parameters given by,
\begin{eqnarray}
\alpha^{\ast} &=& \alpha+x^+,\label{up1}\\
\beta^{\ast} & = & \beta+t.\label{up2}
\end{eqnarray}

The decision-maker must now calculate the best estimator of the
expected number of $X$  given the past experience. To achieve this, they must calculate the Bayes estimator of $F(\theta)$, which can be attained by updating the parameters provided in \eqref{up1}-\eqref{up2} in $\E(\theta)=\alpha/\beta$ to get
\begin{eqnarray}
F(x^+,t) = \E_{\pi(\theta|\tilde x)}(F(\theta)|\tilde x) = \frac{\alpha+x^+}{\beta+t}.\label{bt}
\end{eqnarray}

We now have all the ingredients needed to formulate an expression for the impact score to a journal which $x^+$ cites in the previous $t$ years. First, we normalize the expression given in \eqref{bt} such that at the outset $(x^+,t)=(0,0)$, which is given by and due that the impact score should be revised every two years, we allow $t$ to take even values including the zero. That is,
\begin{eqnarray}
F^{\ast}(x^+,t)=\frac{F(x^+,c)}{F(0,0)}=\frac{\beta}{\alpha} \frac{\alpha+x^+}{\beta+t},\quad x^+=0,1,\dots; t=0,2,4,\dots \label{taxp1}
\end{eqnarray}

The expression \eqref{taxp1} assures us that at the beginning of the process, i.e. $(x^+, t) = (0,0)$, the impact score is 1. This is the impact score for a new journal which enters into the system, since no available information exists for it. The remaining combinations of $x^+$ and $t$ are expressed as a percentage of this first value.

In practice, if the decision-maker wishes the initial rate to be a value other than 1, it could simply be replaced \eqref{taxp1} by
\begin{eqnarray}
F^{\ast}(x^+,t)=\frac{\omega\,F(x^+,c)}{F(0,0)}=\frac{\omega\,\beta}{\alpha} \frac{\alpha+x^+}{\beta+t},\quad x^+=0,1,\dots; t=0,2,4,\dots \label{factor}
\end{eqnarray}
where $\omega>0$ is a constant.

\subsection{Properties of the Bayesian Impact Score and its elasticity}
The following relations of the Bayes rate suggested are compatible with the ideas about a impact score should verify,
\begin{eqnarray}
\Delta F^{\ast}_{x^+}(x^+,t) &=& F^{\ast}(x^++1,t)-F^{\ast}(x^+,t) = \frac{\omega\,\beta}{\alpha(\beta+t)}>0,\label{firstr}\\
\Delta F^{\ast}_{t}(x^+,t) &=& F^{\ast}(x^+,t+2)-F^{\ast}(x^+,t) =-\frac{2\omega\beta(\alpha+x^+)}{\alpha(\beta+t)(\beta+t+2)}<0.\label{secondr}
\end{eqnarray}

The equation \eqref{firstr} suggests that, all else being equal, the impact score will increase as the number of citations increases. In contrast, according to equation \eqref{secondr}, the impact score will decrease as $t$ increases, while keeping the number of citations constant.

Observe that it is possible to write \eqref{factor} as
\begin{eqnarray}
F^{\ast}(x^+,t)=\frac{\omega\,\beta}{\alpha}\left[\gamma(t) \bar x+\left[1-\gamma(t)\right]\E(\theta)\right],\quad 0<\gamma(t)<1,\label{be}
\end{eqnarray}
where $\bar x$ is the sample mean and $\gamma(t)=t/(\beta+t)$. That is, this Bayesian estimate of the impact score can be written as the convex (weighted) sum of the prior mean of $\Theta$, and the mean of the number of cites in the sample, where the weighted factor $\gamma(t)$ can be rewritten \citetext{see \citealp{ericson_1969}} as
\begin{eqnarray*}
\gamma(t) = \frac{t}{\beta+t}=\frac{t\; var\left[\E(X|\theta)\right]}{t\; var\left[\E(X|\theta)\right]+\E\left[var(X|\theta)\right]}.
\end{eqnarray*}

The quantity $var[\E(X|\theta)]$ provides a measure of the heterogeneity of subject categories, while $\E[var(X|\theta)]$ represents a global measure of the dispersion of these means across all journals in the population. It should be noted that $\gamma(t)$ is a specific function that depends on the model parameters, such as $\beta$ in this case, rather than an arbitrary constant, and is determined by both the journal and the collective citation experiences.

Moreover, it is worth mentioning that $\gamma(t)$ is supported on the interval $(0,1)$ and approaches 0 as $\beta\to 0$, while $\gamma(\infty)\to 1$, although these are only limiting cases. In practice, the value of $\gamma(t)$ falls between 0 and 1, implying that $F^{\ast}(x^+,t)$ will always depend on the estimated parameters of the model, which will be obtained using an empirical Bayes approach. However, these extreme values have meaningful interpretations: when $\gamma(t)$ is close to 1, the journal's prior experience has a more significant impact on the final impact score, whereas when $\gamma(t)=0$, less weight is assigned to this experience, and the a priori information plays a more dominant role.

Table \ref{tab3} presents the impact score values, as computed using the formula in \eqref{factor}, for various journals and different values of $x^+$ and $t$, based on the parameter estimates provided in Table \ref{tab2}.

In addition, expressions \eqref{factor}, \eqref{firstr}, and \eqref{secondr} can be employed to estimate the elasticity of the impact score, which measures the impact of changes in the number of citations on the score for a given journal or subject category. This estimator is obtained by
\begin{eqnarray}
\eta_{x^+} = \frac{\Delta F_{x^+}^{\ast}(x^+,t)}{\Delta x^+}\frac{x^+}{F^{\ast}(x^+,t)}=\frac{x^+}{\alpha+x^+}.\label{elasx}
\end{eqnarray}

Observe that from \eqref{elasx}, we have that $\eta_{x^+}<1$, i.e. an inelastic function, being an increasing function on $x^+$ and decreasing with respect to $\alpha$. Then that category (journal) with an estimated value of $\alpha$ greater than another category (journal) will have a worse response in the posterior mean of $X$ from one period to another.

On the other hand, the elasticity of \eqref{be} with respect to $t$ results
\begin{eqnarray}
\eta_{t} = \frac{\Delta F_{t}^{\ast}(x^+,t)}{\Delta t}\frac{t}{F^{\ast}(x^+,t)}=-\frac{t}{\beta+t+2},\label{elast}
\end{eqnarray}
which results negative and again lower than 1. Furthermore, it is an increasing function on $\beta$.

\section{Numerical experiments}
Table \ref{tab2} displays the parameter estimates for the journals using both Poisson and negative binomial distributions. Model selection was based on Akaike's information criterion, which is calculated as $\mbox{AIC}=2(k-\ell_{\max})$, where $k$ represents the number of model parameters and $\ell_{\max}$ represents the maximum value of the log-likelihood function. For further details on this criterion, see \cite{akaike1974}. A lower value of AIC is indicative of a better model fit, and thus desirable for model selection.

\begin{table}[htbp]
\caption{Estimates, via maximum likelihood, of the parameters of the Poisson and negative binomial distributions and AIC values for the different journals studied\label{tab2}}
\begin{center}
\begin{tabular}{lrrrrrr}\hline
Journal & \multicolumn{2}{c}{Poisson} && \multicolumn{3}{c}{NB} \\ \cline{2-3}\cline{5-7}
 & $\widehat\theta$ & AIC && $\widehat\alpha$ & $\widehat\beta$ & AIC \\ \cline{2-3}\cline{5-7}
Ag.Cell                 &  9.21 & 61920.80   &&   1.99 & 0.21 & 42425.50 \\
Amer. Econ. Rev.        & 13.68 & 71880.30   &&   1.03 & 0.07 & 29571.10 \\
IEEE Comm. Mag.         &  4.68 & 21884.30   &&   1.27 & 0.27 & 13981.00 \\
Acc. Chem. Res.         &  8.59 & 219046.00  &&   1.08 & 0.12 & 112400.00 \\
BMC Med.                &  8.22  & 83089.30  &&   1.45 & 0.17 & 51546.40  \\
Adv. in Theor. \& MP    &  17.58 & 28589.80  &&   1.34 & 0.07 & 12752.40 \\
Am. Ant.                & 21.67  & 40815.70  &&   1.15 & 0.05 & 14954.80    \\
IEEE Trans. Inf. Th.    & 14.64 & 172890.00  &&  1.14  & 0.07 & 72734.40 \\
 \hline

\end{tabular}
\end{center}
\end{table}

\begin{figure}[htbp]
\begin{center}
\includegraphics[page=11,width=0.35\linewidth]{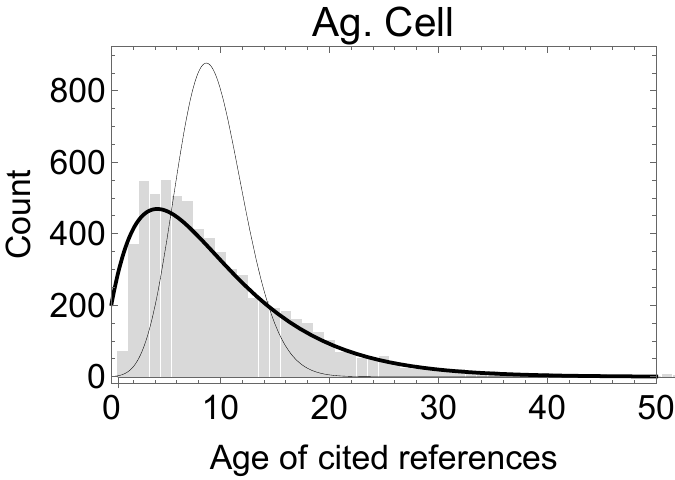}\hspace{0.25cm}
\includegraphics[page=12,width=0.35\linewidth]{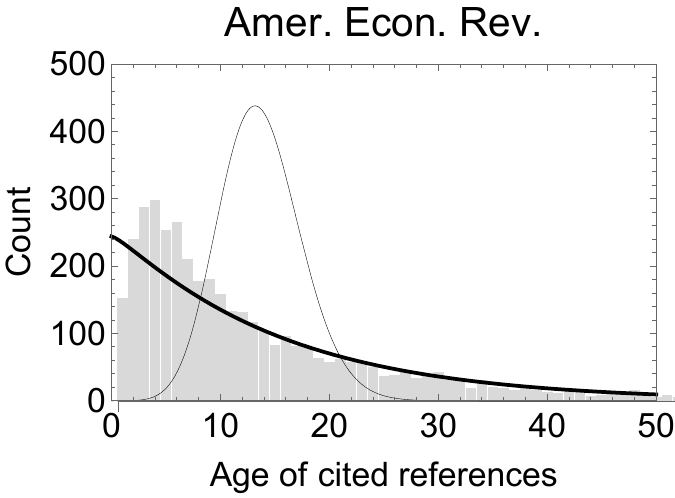}\\
\includegraphics[page=13,width=0.35\linewidth]{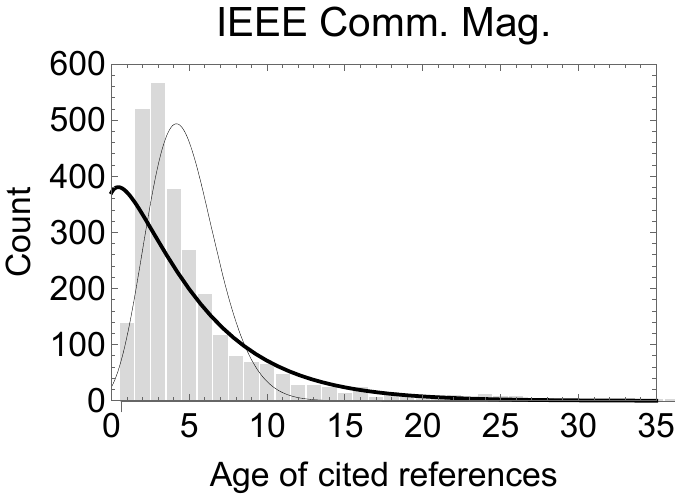}\hspace{0.25cm}
\includegraphics[page=14,width=0.35\linewidth]{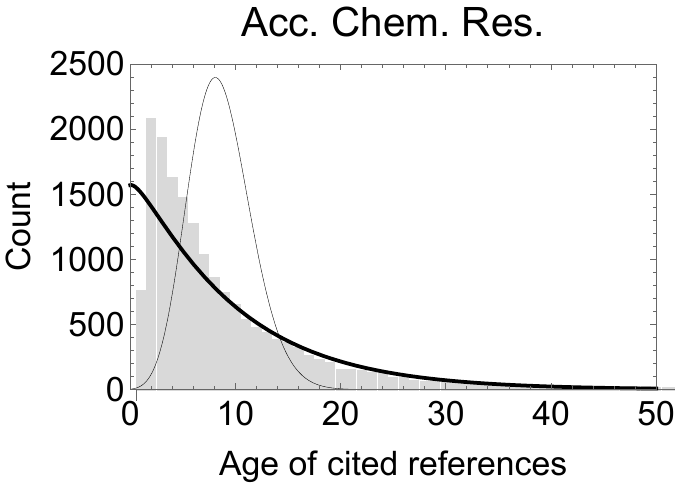}\\
\includegraphics[page=15,width=0.35\linewidth]{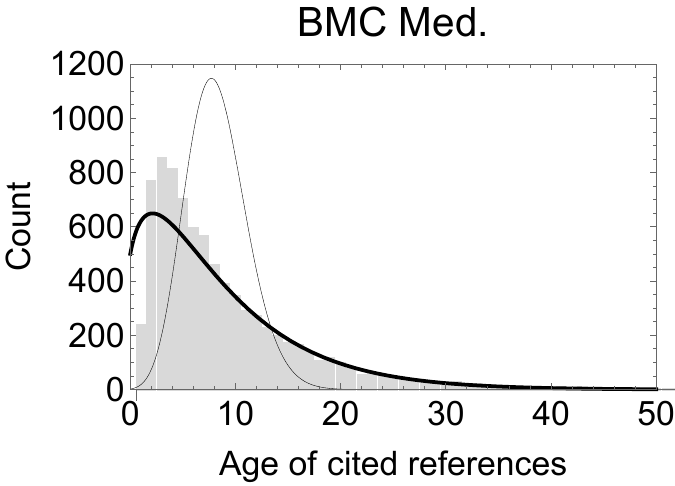}\hspace{0.25cm}
\includegraphics[page=16,width=0.35\linewidth]{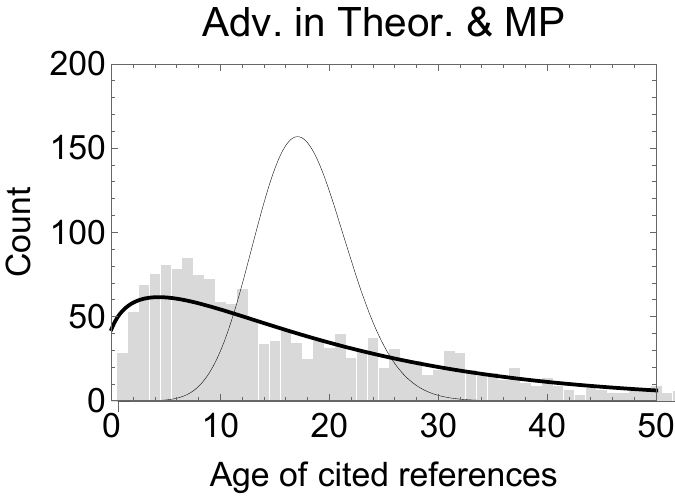}\\
\includegraphics[page=17,width=0.35\linewidth]{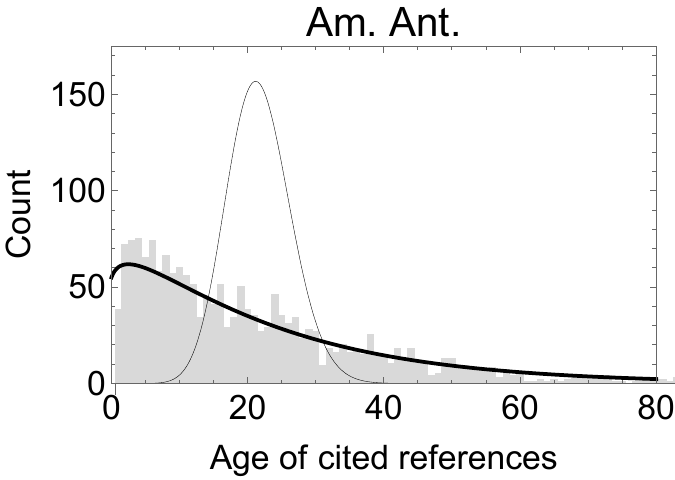}\hspace{0.25cm}
\includegraphics[page=18,width=0.35\linewidth]{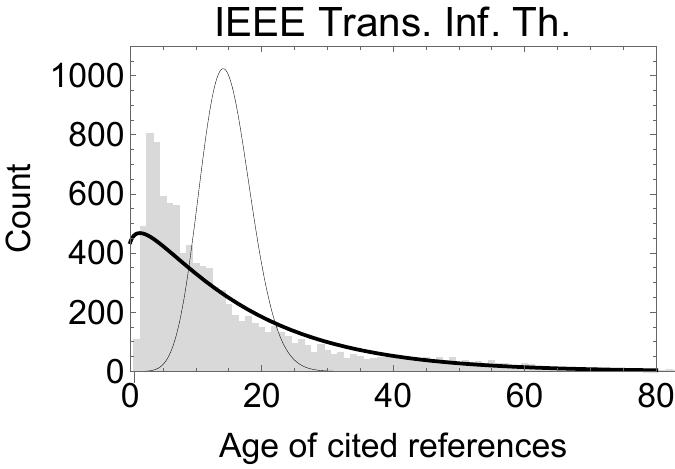}\\
\caption{Empirical and fitted distribution using Poisson (thin line) and negative binomial (thick line) distributions\label{fig1}}
\end{center}
\end{figure}

In all cases, it is apparent that the negative binomial distribution outperforms the Poisson distribution for the datasets considered. This superiority of the negative binomial distribution over the Poisson distribution and the empirical data is further supported by Figure \ref{fig1}.

\begin{figure}[htbp]
\begin{center}
\includegraphics[page=9,width=0.65\linewidth]{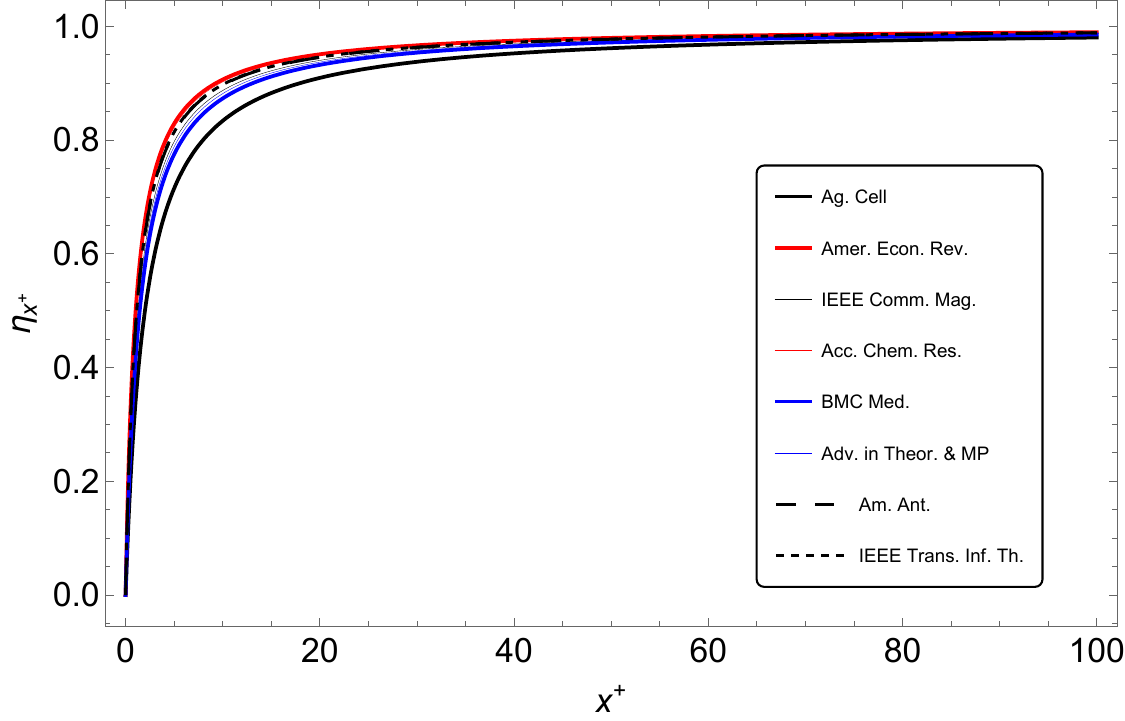}\\
\bigskip
\includegraphics[page=10,width=0.65\linewidth]{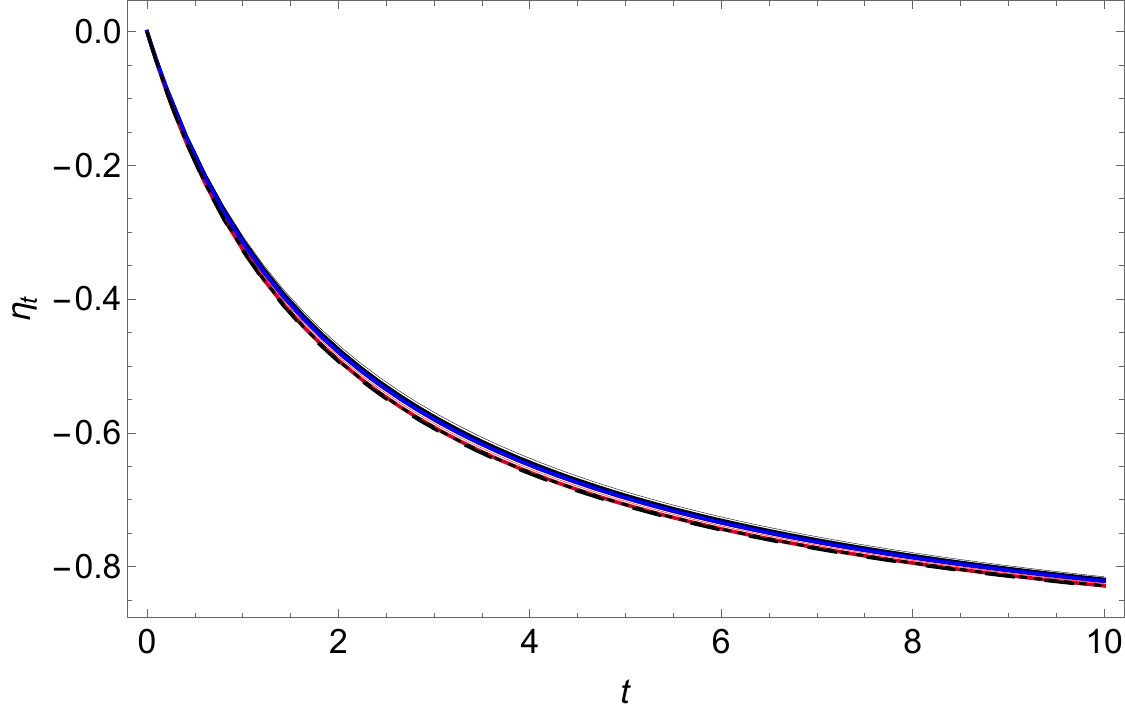}
\caption{Elasticity of the impact score for the different journals considered}\label{fig2}
\end{center}
\end{figure}

Figure \ref{fig2} shows the graphics of elasticities given in \eqref{elasx} and \eqref{elast} for the different journals considered. The citation elasticity (time elasticity) of the impact score is a measure used to show the degree of response, or elasticity, of a journal's impact to changes in the number of citations received (time). It gives the percentage change in the impact generated in relation to a unit percentage change in the citations (time), considering that time (citations) remain constant (ceteris paribus).

In general, the impact score of a journal can be considered inelastic (or relatively inelastic) when the elasticity is less than one (in its absolute value); this happens when changes in citations (time) have a relatively small effect on the impact generated by the journal. As seen in Figure \ref{fig2}, the impact score of the journals considered in this application is inelastic with respect to citations and time (in absolute value), although its elasticity tends to increase with the volume of citations and time.

The elasticity is therefore a measure of the sensitivity (or responsiveness) of the impact score to changes in its variables. The formula returns a positive or negative result, depending on whether the relationship between the variable and the impact is direct or inverse. For example, if citation increases by 10\% and impact increases by 5\%, the citation elasticity is 5\%/10\% = 0.5. However, if time (the age of the cited reference) increases by 10\% and impact decreases by 5\%, the time elasticity is --5\%/10\% = --0.5.

It should be noted that the conclusions drawn here are based on a small sample of journals and not on the entire Scopus publication corpus. As can be seen in Figure \ref{fig2}, the differences between the journals considered in this empirical application in terms of elasticities are small, almost imperceptible to the naked eye in the case of time elasticity. The elasticity of the impact score with respect to the number of citations (citation elasticity) is higher for journals in the social sciences and humanities (i.e. economics and archaeology) and lower for journals in the natural sciences and health (i.e. biology and medicine). The opposite relationships are observed for the elasticity of the impact score with respect to time (time elasticity) in absolute value. Except for one engineering journal (IEEE Comm. Mag.) that exhibits the smallest absolute value of time elasticity, the impact score's elasticity in relation to time is less for social sciences and humanities journals (such as economics and archaeology) and greater for natural sciences and health journals (such as biology and medicine). Note that in the case of negative values, the smallest numbers become the largest in absolute value.

\begin{table}[htbp]
  \resizebox{0.85\textwidth}{!}{\begin{minipage}{\textwidth}
       \caption{Estimated factor for the journals considered assuming that $\omega = 1$.}\label{tab3}
\begin{center}
\begin{tabular}{rcccccccccccl}\hline \hline
 & \multicolumn{11}{c}{Number of cites: $x^+$}\\ \cline{2-12}
Year: $t$ & 0 & 10 &  20 &  30 &  40 &  50 &  60 &  70 &  80 &  90 &  100 \\ \hline

2 &  0.10 & 0.57 & 1.05 & 1.53 & 2.01 & 2.48 & 2.96 & 3.44 & 3.92 & 4.39 & 4.87 & Ag. Cell\\
4 &  0.05 & 0.30 & 0.55 & 0.80 & 1.05 & 1.30 & 1.55 & 1.80 & 2.06 & 2.31 & 2.56 \\
6 &  0.03 & 0.20 & 0.37 & 0.54 & 0.71 & 0.88 & 1.05 & 1.22 & 1.39 & 1.56 & 1.73 \\
8 &  0.03 & 0.15 & 0.28 & 0.41 & 0.54 & 0.67 & 0.80 & 0.93 & 1.05 & 1.18 & 1.31 \\
\hline

2 &  0.03 & 0.36 & 0.69 & 1.02 & 1.35 & 1.68 & 2.00 & 2.33 & 2.66 & 2.99 & 3.32 & Amer. Econ. Rev.\\
4 &  0.02 & 0.18 & 0.35 & 0.52 & 0.69 & 0.85 & 1.02 & 1.19 & 1.35 & 1.52 & 1.69 \\
6 &  0.01 & 0.12 & 0.24 & 0.35 & 0.46 & 0.57 & 0.68 & 0.80 & 0.91 & 1.02 & 1.13 \\
8 &  0.00 & 0.09 & 0.18 & 0.26 & 0.35 & 0.43 & 0.51 & 0.60 & 0.68 & 0.77 & 0.85 \\
\hline

2 &  0.12 & 1.06 & 1.99 & 2.93 & 3.87 & 4.80 & 5.74 & 6.67 & 7.61 & 8.55 & 9.48 & IEEE Comm. Mag.\\
4 &  0.06 & 0.56 & 1.06 & 1.56 & 2.05 & 2.55 & 3.05 & 3.55 & 4.05 & 4.54 & 5.04 \\
6 &  0.04 & 0.38 & 0.72 & 1.06 & 1.40 & 1.74 & 2.08 & 2.42 & 2.76 & 3.09 & 3.43 \\
8 &  0.03 & 0.29 & 0.55 & 0.80 & 1.06 & 1.32 & 1.58 & 1.83 & 2.09 & 2.35 & 2.60 \\
\hline

2 &   0.06 & 0.58 & 1.10 & 1.63 & 2.15 & 2.68 & 3.20 & 3.73 & 4.25 & 4.77 & 5.30 & Acc. Chem. Res.\\
4 &   0.03 & 0.30 & 0.57 & 0.84 & 1.11 & 1.38 & 1.65 & 1.92 & 2.19 & 2.46 & 2.73 \\
6 &   0.02 & 0.20 & 0.38 & 0.56 & 0.75 & 0.93 & 1.11 & 1.29 & 1.47 & 1.65 & 1.84 \\
8 &   0.01 & 0.15 & 0.29 & 0.43 & 0.56 & 0.70 & 0.84 & 0.97 & 1.11 & 1.25 & 1.38 \\
\hline

2 &  0.08 & 0.62 & 1.16 & 1.70 & 2.24 & 2.78 & 3.32 & 3.86 & 4.40 & 4.94 & 5.48 & BMC Med.\\
4 &   0.04 & 0.32 & 0.60 & 0.88 & 1.17 & 1.45 & 1.73 & 2.01 & 2.29 & 2.57 & 2.85 \\
6 &   0.03 & 0.22 & 0.41 & 0.60 & 0.79 & 0.98 & 1.17 & 1.36 & 1.55 & 1.74 & 1.93 \\
8 &   0.02 & 0.16 & 0.31 & 0.45 & 0.59 & 0.74 & 0.88 & 1.03 & 1.17 & 1.31 & 1.46 \\
\hline

2 &  0.03 & 0.29 & 0.54 & 0.79 & 1.04 & 1.30 & 1.55 & 1.80 & 2.05 & 2.31 & 2.56 & Adv. in Theor. \& MP\\
4 &   0.02 & 0.15 & 0.27 & 0.40 & 0.53 & 0.66 & 0.79 & 0.92 & 1.04 & 1.17 & 1.30 \\
6 &   0.01 & 0.10 & 0.18 & 0.27 & 0.36 & 0.44 & 0.53 & 0.61 & 0.70 & 0.79 & 0.87 \\
8 &   0.00 & 0.07 & 0.14 & 0.20 & 0.27 & 0.33 & 0.40 & 0.46 & 0.53 & 0.59 & 0.66 \\
\hline

2 &  0.02 & 0.24 & 0.45 & 0.66 & 0.87 & 1.08 & 1.30 & 1.51 & 1.72 & 1.93 & 2.15 & Am. Ant.\\
4 &   0.01 & 0.12 & 0.23 & 0.33 & 0.44 & 0.55 & 0.66 & 0.76 & 0.87 & 0.98 & 1.09 \\
6 &   0.00 & 0.08 & 0.15 & 0.22 & 0.30 & 0.37 & 0.44 & 0.51 & 0.58 & 0.66 & 0.73 \\
8 &   0.00 & 0.06 & 0.11 & 0.17 & 0.22 & 0.28 & 0.33 & 0.38 & 0.44 & 0.49 & 0.55 \\
\hline

2 &   0.03 & 0.33 & 0.63 & 0.92 & 1.22 & 1.52 & 1.81 & 2.11 & 2.41 & 2.70 & 3.00 & IEEE Trans. Inf. Th.\\
4 &   0.02 & 0.17 & 0.32 & 0.47 & 0.62 & 0.77 & 0.92 & 1.07 & 1.22 & 1.38 & 1.53 \\
6 &   0.01 & 0.11 & 0.21 & 0.32 & 0.42 & 0.52 & 0.62 & 0.72 & 0.82 & 0.92 & 1.02 \\
8 &   0.00 & 0.08 & 0.16 & 0.24 & 0.31 & 0.39 & 0.47 & 0.54 & 0.62 & 0.69 & 0.77 \\
\hline\hline

\end{tabular}
\end{center}
\end{minipage}}
\end{table}

In the interpretation at the paper level, the impact score of the paper appears directly in Table \ref{tab3} when crossing the age of the paper ($t$) and the citations received ($x^+$). This normalization of the impact of a paper is in relation to the citation frequencies observed in the past in each journal. It is adjusted, therefore, to the obsolescence of the literature on the topic of the journal. For reasons of space, the scores in Table \ref{tab3} are shown with only two decimals and in intervals or classes of two years and 10 citations. However, it can be extended in a similar way to smaller intervals, and even to one citation and one year.

Thus, for example, in the American Economic Review, a highly prestigious publication in economics, a paper with 10 citations after two years (an average of 5 cites/year), its impact can be quantified in a score of 0.36. If said paper does not receive citations in the two subsequent years (third and fourth of life), its score is reduced by half (0.18). On the other hand, if during the third and fourth year of life it receives another 10 additional citations (same average of 5 cites/year), then its score would be 0.35, slightly lower than it was in the second year.

Observe in this example that with a uniform citation distribution of 5 cites/year, the marginal score of the first citations is somewhat higher than that of the last ones. This is because in this journal the articles usually receive fewer citations during the first two years of life than during the following two, so the marginal score of citations in the field defined by the topic of the journal decreases during the first years of life. However, the opposite happens when we exceed the peak of the citation distribution, which for this journal is between 4 and 6 years. This is observed in the rest of the values of the diagonal that are again 0.35.

Now suppose that a different article in this journal receives 20 citations during its first two years (at a rate of 10 cites/year). In this case, its impact score is 0.69. If this citation rate remains constant, then its score remains at 0.69 during the second interval (third and fourth year of life), although it slightly decreases to 0.68 in the following intervals.

As can be seen, the same number of citations has a different impact score depending on the expected frequencies in the topic of each journal. These scores are comparable within the same journal, but also between journals from different fields. Thus, for example, for similar values of citations and years, the scores in the communication journal are much higher (almost double) than those in the chemistry journal, for example. Moreover, scores in medicine are somewhat higher than those in biology. The lowest scores are reached in archaeology and physics, thus indicating that the citation density is higher in these fields.

To empirically analyze the behavior of our Bayesian Score, we used the Dimensions AI database and its article-level normalized impact indicator, the Field Citation Ratio (FCR). The FCR serves as a metric to measure the relative citation performance of a publication compared to other articles within its field and of similar age. An FCR greater than 1 indicates an above-average citation impact as papers in the same field and publication year. For the eight journals listed in Table \ref{tab1}, we randomly selected a simple sample of size N=5048 research articles from the 2014-2021 cohort. This represents approximately 32\% of the total articles published during these years.

The Pearson linear correlation coefficient between the FCR and the Bayesian Score is 0.68. This value indicates a significant correlation between the two measures, as would be expected when trying to measure the same phenomena. However, it also indicates that there are some differences between these normalized measures. We have therefore looked at the descriptive statistics of the data distributions to try to identify where these observed differences between the two indicators come from.

The coefficient of variation (CV) in Table \ref{tab4} is a measure of relative variability used to express the dispersion of a sample relative to its mean. It is calculated by dividing the sample standard deviation (SD) by the mean. This coefficient is useful for comparing the relative dispersion between two data sets, especially when the scales of measurement are different, as is the case here, as the CV is dimensionless. A low CV indicates low variability relative to the mean, while a high CV indicates higher relative variability.

As can be seen in Table \ref{tab4}, the differences in CV between journals from different disciplines are considerable. In the case of the Bayesian Score, these differences are comparatively smaller, indicating greater consistency in this new indicator. This can be seen in the range of variation for this measure across journals (0.62-3.26 for the FCR and 0.67-2.31 for the Bayesian Score). However, the discipline of the journal also appears to have a significant impact on the CV, highlighting the importance of considering the scientific area of application when using either of these indicators. It can also be concluded that the ranking of journals from highest to lowest CV is similar according to both metrics. The only difference is between the American Economic Review and Aging Cell, where they exchange positions between sixth and seventh place depending on the metric considered.

\begin{sidewaystable}
\caption{Descriptive statistics for Field Citation Ratio (FCR) and Bayesian Score in a random sample. The coefficient of variation (CV) is defined as the ratio of the standard deviation (SD) to the mean (CV = SD/mean)}\label{tab4}
\begin{center}
\begin{tabular}{lcccccccccc}
\toprule
{FCR} & $n$& {Mean} & {SE} & {Median} & {SD} & {CV} & {Kurtosis} & {Skewness} & $\min$ & $\max$ \\
\midrule
{Ag. Cell} & 385 & 6.53 & 0.298 & 4.32 & 5.847 & 0.90 & 3.453 & 1.766 & 0 & 35.5 \\
{Amer. Econ. Rev.} & 346 & 18.52 & 0.719 & 16.38 & 13.368 & 0.72 & 0.086 & 0.748 & 0.7 & 67.7 \\
{IEEE Comm. Mag.} & 1072 & 4.88 & 0.253 & 1.24 & 8.290 & 1.70 & 5.880 & 2.390 & 0 & 52.4 \\
{Acc. Chem. Res.} & 454 & 7.61 & 0.222 & 6.94 & 4.739 & 0.62 & 0.597 & 0.811 & 0.3 & 24.9 \\
{BMC Med.} & 597 & 8.74 & 0.339 & 5.82 & 8.286 & 0.95 & 4.209 & 1.732 & 0 & 56.3 \\
{Adv. in Theor. \& MP} & 124 & 3.43 & 0.521 & 1.73 & 5.797 & 1.69 & 36.923 & 5.294 & 0 & 49.9 \\
{Am. Ant.} & 439 & 3.13 & 0.487 & 0.00 & 10.204 & 3.26 & 116.771 & 9.602 & 0 & 144.9 \\
{IEEE Trans. Inf. Th.} & 1631 & 4.14 & 0.127 & 2.57 & 5.124 & 1.24 & 11.326 & 3.043 & 0 & 44.1 \\
\bottomrule

{Bayesian Score} & $n$ & {Mean} & {SE} & Median & {SD} & {CV} & {Kurtosis} & {Skewness} & $\min$ & $\max$ \\
\midrule
{Ag. Cell} & 385 & 0.94 & 0.036 & 0.57 & 0.702 & 0.75 & 4.099 & 1.708 & 0 & 4.9 \\
{Amer. Econ. Rev.} & 346 & 0.69 & 0.031 & 0.52 & 0.581 & 0.85 & 3.003 & 1.532 & 0 & 3.3 \\
{IEEE Comm. Mag.} & 1072 & 0.75 & 0.039 & 0.12 & 1.284 & 1.70 & 10.709 & 2.868 & 0 & 9.5 \\
{Acc. Chem. Res.} & 454 & 1.58 & 0.049 & 1.38 & 1.053 & 0.67 & 1.630 & 1.152 & 0 & 5.3 \\
{BMC Med.} & 597 & 0.90 & 0.035 & 0.62 & 0.853 & 0.94 & 4.427 & 1.770 & 0 & 5.5 \\
{Adv. in Theor. \& MP} & 124 & 0.08 & 0.011 & 0.03 & 0.121 & 1.61 & 19.965 & 3.996 & 0 & 0.9 \\
{Am. Ant.} & 439 & 0.05 & 0.005 & 0.02 & 0.111 & 2.31 & 45.134 & 5.592 & 0 & 1.3 \\
{IEEE Trans. Inf. Th.} & 1631 & 0.17 & 0.006 & 0.11 & 0.223 & 1.33 & 17.910 & 3.322 & 0 & 2.7 \\
\bottomrule
\end{tabular}
\end{center}
\end{sidewaystable}

The CV is therefore related to the discriminatory power of a measure. The discriminatory power of an impact measure refers to its ability to effectively discriminate between the impact levels of different papers. In the context of citation-based impact measures, the CV plays a role in this discrimination. A higher CV implies greater variability in the citation scores of different publications, suggesting a wider range of impact levels. This can increase the discriminatory power, as it reflects more pronounced differences in the impact of individual papers. Researchers or evaluators may find such a measure valuable in identifying and prioritizing highly impactful papers, taking into account a more diverse impact landscape. Conversely, a lower CV implies a more uniform distribution of citations, which may lead to reduced discriminatory power. While this may provide a more consistent view of impact across papers, it may struggle to discriminate subtle differences in impact. The discriminatory power of an impact measure therefore depends on the specific aims and preferences of the analysis. A higher CV might be preferred if the aim is to capture different levels of impact, while a lower CV might be chosen for a more uniform and stable assessment of impact.

In essence, a higher CV highlight the heterogeneity in the impact of scholarly papers and emphasize that a few can dominate the citation landscape. Researchers and evaluators considering impact measures need to be aware of these dynamics and choose metrics that are consistent with their objectives, whether they seek to capture diversity in impact (higher CV) or prioritize stability and uniformity (lower CV).

\section{Conclusions}
A central problem in bibliometrics is how to compare the impact of publications across different fields and years. While there is a large literature on field-normalized citation counts, there is much less research on time normalization.

Traditionally, citation counts are divided by the number of years since a publication's release to simplify the comparison process. This approach assumes that all citations are equally valuable, regardless of when they were made. However, as research publications age, their usage tends to decrease, and authors tend to cite more recent publications instead of older ones, resulting in a phenomenon known as literature obsolescence.

Furthermore, older papers have had more time to accumulate citations, but due to the increasing rate at which the scientific corpus is expanding, older influential papers had a lower citation potential shortly after publication than younger influential papers. These issues highlight the importance of developing a time normalization method that considers these factors for more accurate impact comparisons.

Both because of the phenomenon of obsolescence, which causes the density of citations to decrease as the age of the cited document increases, and because of the increasing rate of growth of the scientific corpus, citations to older documents should be more widely recognised than citations to more recent documents.

The empirical citation data show heterogeneity in the distributions. Therefore, the Bayesian approach is used to solve the problem of field and time normalization. In this approach, the a priori distribution of citations in each field is used to estimate the expected a posteriori distribution in that field. This a posteriori distribution is then used to normalize the citations received by a publication in that field.

The proposed Bayesian Impact Score is increasing and concave with the number of citations received and decreasing and convex with the age of the publication. This means that the marginal score of an additional citation decreases with the cumulative number of citations and increases with the time since publication.

This impact score gives less value to citations received in the first few years after publication and reduces its value as the accumulated volume of citations increases. This reduction in score with citation volume attempts to mitigate the effect whereby highly cited documents attract more attention and tend to receive more citations simply because they are highly cited.

Some considerations can be made. The well-known Journal Impact Factor favours journals that concentrate a large proportion of their citations in the first few years after publication, i.e. journals in fields with high obsolescence. However, our proposal, which gives an increasing marginal score over time, is fairer in fields with less obsolescence.

A citation loss function is used to fit the citations received by a journal to the expected citation distribution in its field. It is a bivariate measure and its properties include that it increases with citations, decreases over time, and is comparable across fields. It is also aggregable in the sense that it can be used at the level of authors by simply adding the scores of each of their publications.

In conclusion, the normalization of citation impact indicators is crucial for accurate assessment of research output, and this can be achieved by assigning publications to specific fields. Although the WoS journal subject categories are still the most commonly utilized field classification system, questions have emerged about the reliability of normalized indicators based on the selection of the classification system. In addition, scholars have investigated the feasibility of substituting alternative classification systems for the WoS journal subject categories. It is important to note that our proposal avoids using a journal classification system altogether, instead utilizing the journal of publication to define expected citations within the subject area of each paper. This approach addresses some of the concerns raised about the use of traditional classification systems, and could potentially offer a more robust and accurate method for normalization.

\subsubsection*{Funding and/or Conflicts of interests/Competing interests:}
The authors have no competing interests to declare that are relevant to the content of this article.

\subsubsection*{Acknowledgement}The authors would like to thank to the Editor and two anonymous referees for valuable
comments and suggestions which improved the presentation of the paper.

\end{document}